\documentclass[english,letterpaper]{article}
\usepackage[T1]{fontenc}
\usepackage[latin1]{inputenc}
\usepackage{graphicx}

\makeatletter

\input{epsf}
\makeatletter

\usepackage{geometry}

\makeatother

\usepackage{babel}
\makeatother
\begin{document}

\title{Thermal fluctuations propagation in the relativistic Euler regime:
a causal appraisal}

\author{A. Sandoval-Villalbazo and D. Brun \\
 Universidad Iberoamericana \\
 Prolongaci\'{o}n Paseo de la Reforma 880 \\
 M\'{e}xico, D.F. 01219, M\'{e}xico}

\maketitle
\begin{abstract}
It is shown that thermal fluctuations present in a simple non-degenerate
relativistic fluid satisfy a wave equation in the Euler regime. The
characteristic propagation speeds are calculated and the classical
expression for the speed of sound is recovered at the non-relativistic
limit. Implications and generalizations of this work are analyzed. 
\end{abstract}

\section{Introduction}

Relativistic transport theory has dramatically increased its interest
due to the detection of high temperature plasmas produced in the Relativistic
Heavy Ion Collider (RHIC). In this context, an Euler fluid description
provides a good approximation for events involving Au-Au collisions
\cite{rev1}. Generalizations involving dissipative effects have been
developed taking into account recent experimental data \cite{Chau1}.
The first works regarding relativistic hydrodynamics can be tracked
down to the pioneering 1940 Eckart's monographs \cite{Eckart}, and
to the relativistic fluids section included in the Landau-Lifshitz
fluid mechanics textbook \cite{LL}. The explicit form of the linearized
transport equations obtained within Eckart's framework raised serious
doubts concerning the stability and causality properties of the system
\cite{HL} \cite{Israel1}. Indeed, it was only recently observed
that the so-called stability problem in relativistic hydrodynamics
is due to the heat-acceleration coupling introduced in Eckart's work
\cite{GRG1}. Following these ideas, it became pertinent to revise
the causality properties in this kind of systems, focusing in the
possibility of generating a hyperbolic partial differential equation
describing \textit{temperature} fluctuations. This work tackles the
problem for an Euler fluid and suggests some new insights for this
issue while examining the linearized equations in the Navier-Stokes
regime without resorting to extended formalisms.

In section 2 the basic formalism is presented on the basis of relativistic
kinetic theory for an inert dilute fluid, emphasizing the role of
the Enskog transport equation. Section 3 is devoted to the analysis
of the linearized transport equations in the Euler regime, by means
of a derivation of a wave equation describing thermal fluctuations
in the relativistic case, and its corresponding non-relativistic limit.
Some final thoughts regarding the causal properties of relativistic
fluids in the dissipative case are included in the final section of
this work.

\section{Kinetic foundations and transport equations}

\bigskip{}

For decades, relativistic kinetic theory has been successfully applied
in the study of high temperature fluids \cite{Cer}. The starting
point here is the relativistic Boltzmann equation for a simple fluid
in the absence of external forces:

\bigskip{}
 \begin{equation}
v^{\alpha}f_{,\alpha}=J(ff^{\prime})\label{B1}\end{equation}

In Eq. (\ref{B1}), $f$ is the distribution function in the phase
space, $J(ff^{\prime})$ is the collisional kernel, and the molecular
four velocity, $v^{\alpha}$ is given by \begin{equation}
v^{\alpha}=(\gamma w^{l},\gamma c)\end{equation}
 where $w^{l}$ is the molecular velocity (three spatial components).
As usual, $\gamma=\left(1-\frac{w^{l}w_{l}}{c^{2}}\right)^{-1/2}$.
All latin indices run from 1 to 3 and the greek ones run up to 4.
A signature $(1,1,1,-1)$ is taken, so that $u^{\alpha}u_{\alpha}=-c^{2}$.
The relativistic generalization of Enskog's transport equation can
be casted in the form \cite{Cer} \cite{PA1}: \begin{equation}
\frac{\partial}{\partial t}\left(n\left\langle \Psi\right\rangle \right)+\left(n\left\langle w^{l}\Psi\right\rangle \right)_{;l}=0\end{equation}
 where $n$ is the particle number density and the average of the
collisional invariant $\left\langle \Psi\right\rangle $ is defined
as \begin{equation}
\langle\Psi\rangle=\frac{1}{n}\int\gamma\Psi fdv^{\ast}\label{prom}\end{equation}
 with $dv^{\ast}=$ $\gamma^{5}\frac{cd^{3}w}{v^{4}}$ \cite{Liboff}.

In the Euler regime, all averages are calculated using the equilibrium
(Juttner) distribution function, valid for a non-degenerate gas \cite{Cer}:
\begin{equation}
f^{(0)}=\frac{n}{4\pi c^{3}K_{2}(\frac{1}{z})}e^{\frac{u^{\beta}v_{\beta}}{zc^{2}}}\label{Juttner}\end{equation}
 where $u^{\beta}=\langle v^{\beta}\rangle$ is the hydrodynamic velocity,
$z=\frac{kT}{mc^{2}}$ is the relativistic parameter and $K_{2}$
is the modified Bessel function of the second kind. Derivatives with
respect to $u^{\beta}$ can be explicitly evaluated in Eq.(\ref{Juttner}).
After this operation, for the sake of simplicity, all calculations
will be performed in the comoving frame of the fluid.

Now, the collisional invariants are $\Psi=1$ (a constant), $mw^{l}\gamma$
(the three-momentum) and $mc^{2}\gamma$ (the mechanical energy).
For $\Psi=1$ the continuity equation follows immediately \begin{equation}
\frac{\partial}{\partial t}\left(n\right)+(nu^{l})_{;l}=0\label{cont1}\end{equation}
 Substituting $\Psi=mw^{l}\gamma$, the momentum balance equation
is obtained: \begin{equation}
\frac{\partial}{\partial t}\left(n\left\langle mw^{k}\gamma\right\rangle \right)+\left(nm\left\langle w^{l}w^{k}\gamma\right\rangle \right)_{;l}=0\label{mom1}\end{equation}
 The use of Eqs. (\ref{prom}) and (\ref{Juttner}) allows to rewrite
Eq. (\ref{mom1}) in terms of the local thermodynamic variables: \bigskip{}
 \begin{equation}
\frac{1}{c^{2}}(n\varepsilon+p)\frac{\partial}{\partial t}\left(u^{l}\right)+kn\frac{\partial T}{\partial x^{l}}+kT\frac{\partial n}{\partial x^{l}}=0\label{mom2}\end{equation}
 where the internal energy per particle $\varepsilon$ reads: \begin{equation}
\varepsilon=3nkT+nmc^{2}\frac{K_{1}(1/z)}{K_{2}(1/z)}\label{eint}\end{equation}
 and the pressure satisfies the sate equation \begin{equation}
p=nkT\label{edo}\end{equation}

Finally, for $\Psi=mc^{2}\gamma$, the resulting balance equation
reads:

\begin{equation}
\frac{\partial}{\partial t}\left(n\left\langle \gamma mc^{2}\right\rangle \right)+\left(n\left\langle w^{l}(\gamma mc^{2})\right\rangle \right)_{;l}=0\label{energ1}\end{equation}
 or, in terms of the thermodynamic variables: \begin{equation}
\frac{\partial(n\varepsilon)}{\partial t}+p\theta=0\label{energ2}\end{equation}
 in Eq. (\ref{energ2}) we have defined $\theta=u_{;\alpha}^{\alpha}$.
The set of equations (\ref{cont1},\ref{mom2},\ref{energ2}) is highly
nonlinear and its full treatment is rather complex. For a system close
to equilibrium we shall linearize this set in order to perform a fluctuation
analysis for the thermodynamical variables.

\section{Linearized equations and causality analysis}

In order to proceed with the analysis of the Euler system (\ref{cont1},\ref{mom2},\ref{energ2})
close to equilibrium, we decompose any thermodynamical variable $X$
into a constant average value $X_{o}$ and a space and time dependent
fluctuation $\delta X$, so that \begin{equation}
X=X_{o}+\delta X\label{fluc}\end{equation}
 According to this definition, neglecting second order terms, the
linearized continuity equation, obtained from (\ref{cont1}) reads:
\begin{equation}
\frac{\partial}{\partial t}(\delta n)+n_{o}\,\delta\theta=0\label{cont}\end{equation}
 Analogously, the linearized momentum balance for the longitudinal
mode $\delta\theta$ becomes: \begin{equation}
\tilde{\rho}_{o}\frac{\partial}{\partial t}(\delta\theta)+n_{o}k\nabla^{2}(\delta T)+kT_{o}\nabla^{2}(\delta n)=0\label{momL}\end{equation}
 where we have defined $\tilde{\rho}_{o}=\frac{n_{o}\varepsilon_{o}+p_{o}}{c^{2}}$.
For the linearized energy balance equation we get: \begin{equation}
n_{o}c_{v}\frac{\partial(\delta T)}{\partial t}+n_{o}kT_{o}\delta\theta=0\label{energL}\end{equation}
 Here, the heat capacity (per particle) is given by \begin{equation}
c_{v}=\left(\frac{\partial\varepsilon_{o}}{\partial T}\right)_{n}\end{equation}

In order to decouple the system and establish a partial differential
equation for $\delta T$, we first solve for $\delta\theta$ in both
sides of Eqs. (\ref{cont}) and (\ref{energL}). Equating the results
we obtain the useful relation:

\begin{equation}
\frac{1}{n_{o}}\left(\frac{\partial}{\partial t}\delta n\right)-\frac{c_{v}}{kT_{o}}\left(\frac{\partial}{\partial t}\delta T\right)=0\label{equalT}\end{equation}
 We derive with respect to time in both sides of Eq. (\ref{momL})
: \begin{equation}
\tilde{\rho}_{o}\frac{\partial^{2}}{\partial t^{2}}\delta\theta+n_{o}k\nabla^{2}\frac{\partial}{\partial t}(\delta T)+kT_{o}\nabla^{2}\frac{\partial}{\partial t}(\delta n)=0\end{equation}
 so that, inserting the expression for $\frac{\partial}{\partial t}\delta n$
from equation (\ref{equalT}) we obtain: \begin{equation}
\tilde{\rho}_{o}\frac{\partial^{2}}{\partial t^{2}}\delta\theta+n_{o}k\nabla^{2}\frac{\partial}{\partial t}(\delta T)+kT_{o}\nabla^{2}\left(\frac{n_{o}c_{v}}{kT_{o}}\right)\frac{\partial}{\partial t}(\delta T)=0\label{w1}\end{equation}
 One first time derivative is immaterial in each term of Eq. (\ref{w1}),
so that after an elementary arrangement of terms we can write the
wave equation for thermal fluctuations:

\begin{equation}
\nabla^{2}(\delta T)-\frac{\tilde{\rho}_{o}c_{v}}{kn_{o}T_{o}(c_{v}+k)}\frac{\partial^{2}}{\partial t^{2}}(\delta T)=0\end{equation}
 Thus, the propagation speed ($C_{s}$) of a thermal wave in a relativistic
Euler fluid is: \begin{equation}
C_{s}^{2}=\frac{n_{o}T_{o}(c_{v}+k)k}{\tilde{\rho}_{o}c_{v}}\label{rss}\end{equation}
 In the non-relativistic limit, $z\rightarrow0$, $c_{v}\rightarrow\frac{3}{2}k$
and $\tilde{\rho}_{o}\rightarrow n_{o}m$, so that (\ref{rss}) yields
\begin{equation}
C_{s}^{2}\rightarrow\frac{5}{3}zc^{2}\end{equation}
 which is the non-relativistic speed of sound. Also, the relativistic
propagation speed (\ref{rss}) can be rewritten in terms of $z$ as:
\begin{equation}
C_{s}^{2}=\frac{(c_{v}+k)z}{c_{v}\left(4z+\frac{K_{1}(1/z)}{K_{2}(1/z)}\label{cs}\right)}c^{2}\end{equation}

It is interesting to notice that some authors perform a similar analysis
for $\delta n$ neglecting temperature fluctuations, and only taking
into account Eqs. (\ref{cont1}) and (\ref{mom2}), in order to establish
a wave equation for density fluctuations \cite{Kolb}. In that case
it is immediate to find out that the corresponding propagation speed
is: \begin{equation}
C_{T}^{2}=\frac{z}{\left(4z+\frac{K_{1}(1/z)}{K_{2}(1/z)}\right)}c^{2}\label{ct}\end{equation}

In the same order of ideas, one can make a simple analysis neglecting
the number density fluctuations and taking into account only Eqs.
(\ref{mom2}) and (\ref{energL}). In this case, the expression for
a wave equation for thermal fluctuations reads: \begin{equation}
C_{n}^{2}=\frac{z}{\frac{c_{v}}{k}\left(4z+\frac{K_{1}(1/z)}{K_{2}(1/z)}\right)}c^{2}\label{cn}\end{equation}
 Figure 1 shows a comparison of the characteristic speeds for increasing
$z$.

\begin{figure}

\caption{Comparison of fluctuation propagation speeds for the full relativistic
case(solid), non-relativistic case (long dashed), $\delta n$ fluctuations
neglecting thermal fluctuations (short dashed) and $\delta T$ fluctuations
neglecting number density fluctuations (dotted). \protect\includegraphics{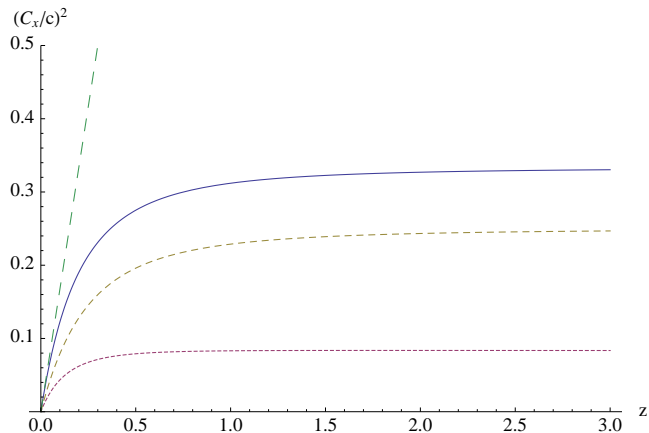}}
\end{figure}

\section{Final remarks}

It has recently been proved the nonexistence of generic instabilities
in the linearized transport equations at the Navier-Stokes regime
\cite{PA1}. In this paper it is shown that, in the Euler regime,
there is no causality problem. The linearized transport equations
become a hyperbolic system and, for further research, it can be taken
as a starting point for a simplified calculation and for validation
of numerical work in the non-linear case.

The non-relativistic limit has been recovered, as expected, and thermal
fluctuations also satisfy a hyperbolic partial differential equation.
In most textbooks, the establishment of the (parabolic) heat equation
is based on an extension of Eq.(\ref{energ2}) including heat conduction,
neglecting velocity fluctuations. On the other hand, if the linearized
equation of motion (\ref{momL}) is taken as the basis of the description
of thermal fluctuations, then a causal equation is obtained for the
non-dissipative fluid. Thus, for the dissipative case it is suggested
that the suitable generalization of the whole linearized system (\ref{cont1},\ref{mom2},\ref{energ2})
should be taken into account, emphasizing the role of Eq.(\ref{mom2})
when analyzing causal properties of the system. \emph{Neglecting velocity
fluctuations clearly leads to non-causality}. It can be noted, also,
that density fluctuations (neglecting the thermal ones) and thermal
fluctuations (neglecting the density ones) present different propagation
speeds, satisfying the relation $C_{n}^{2}+C_{T}^{2}=C_{s}^{2}$.
Moreover, taking $\delta\theta=0$ is unrealistic, since when the
fluid is at rest the mean velocity is zero, but the fluctuations do
not vanish. The approximate expressions (\ref{ct}) and (\ref{cn})
may be useful in particular situations involving dissipative effects.
This opens a line for future research.

The authors wish to thank A.L. Garcia-Perciante for her valuable comments
for this work.

\end{document}